\documentclass[referee]{raa}           
\usepackage{graphicx,times}
\usepackage{natbib}
\usepackage{subfig}
\usepackage{amssymb,amsmath}
\bibpunct{(}{)}{;}{a}{}{,}

\usepackage[CJKbookmarks=true,bookmarksnumbered=true]{hyperref} 
\hypersetup{pdftitle = The title of my PDF, pdfauthor = My name, pdfsubject= The subject, pdfkeywords = keyword1 keyword2 keyword3} 
\hypersetup{colorlinks = true, linkcolor = green, anchorcolor = red, citecolor = blue, filecolor = red, pagecolor = red, urlcolor = red}

\begin{document}
   \title{A geometric delay model for Space VLBI
}

 \volnopage{ {\bf 2012} Vol.\ {\bf X} No. {\bf XX}, 000--000}
   \setcounter{page}{1}

   \author{Shi-Long Liao\inst{1,2}, Zheng-Hong Tang\inst{1}, Zhao-Xiang Qi\inst{1}
   }

   \institute{ Shanghai Astronomical Observatiory, Chinese Academy of Science, Shanghai,200030, 
China; {\it shilongliao@shao.ac.cn}\\
        \and
             University of Chinese Academy of Science,
             Beijing 100049, China\\
\vs \no
   {\small Received 2013 Nov 12; accepted 2013 Dec 27}
}
\abstract{A relativistic delay model for space very long baseline interferometry (hereafter SVLBI) observation of sources at infinite distance is derived. In SVLBI, where one station is on a spacecraft, the orbiting station's maximum speed in an elliptical Earth orbit is much bigger than the ground VLBI (here after GVLBI), leading to a higher delay rate . The delay models inside the VLBI correlators are usually expressed as fifth-order polynomials in time that good for a limited time interval, which are evaluated by the correlator firmware and track the interferometer delays over a limited time interval. The higher SVLBI delay rate requires more accurate polynomial fitting and evaluation, more frequent model updates.
}

   \authorrunning{S.-L. Liao et al. }            
   \titlerunning{A delay model for Space VLBI}  
   \maketitle

%
\numberwithin{equation}{section}
\section{Introduction}           
\label{sect:intro}

The technique of VLBI is widely used in radio astronomical observations. In general, the angular resolution of VLBI is determined by the observing frequency and the length of the baseline. Limited by the size of the Earth, the angular resolution of GVLBI is not good enough for many compact sources. To improve the resolution at a fixed frequency is to place a radio telescope in space, preferably in an elliptical orbit around the Earth. The satellite radio telescope then observes in conjunction with ground-based radio telescopes, synthesizing an aperture whose effective resolution is that of a radio telescope much larger than the Earth. This technique is called SVLBI. SVLBI projects have been proposed or planned by several agencies: Quasat by the European Space Agency and NASA in 1980s (\citealt{Schilizzi+1984,Schilizzi+1988}), VSOP by Institute for Space and Astronautical Science in Japan (\citealt{Hirabayashi+1988}), which was launched successfully in February 12,1997 (\citealt{Hirabayashi+etal+1997a, Hirabayashi+etal+1997b}), and RADIOASTRON by the Space Research Institute of the USSR Academy of Science (\citealt{Kardashev+etal+1987}), which was launched in July 18, 2011 (\citealt{Alexandrov+etal+2012}). The Chinese Space VLBI project (hereafter C-SVLBI), which is proposed by Shanghai Astronomical Observatory and National Space Science Center of China recently, plans to launch \newpage \noindent two antennas into Earth orbit in the first step (\citealt{Hong+etal+2013}; \citealt{Shen+etal+2013}).

Delay model is a representation of the apparent delay in the wave front received at a radio telescope referred to its arrival time at the other telescope. VLBI correlation requires an accurate delay model for each instant of an observation. An accurate delay model corresponds to small residual delays and delay rates in correlation, which enable long coherent integrations in fringe fitting and translate into a capability for detecting fringes on weaker sources.

In GVLBI, the telescope is tied to the Earth's surface, and its acceleration is about $34mm\cdot s^{-2} $. Considering the maximum delay is about 21 $ms$, the effect of acceleration on the baseline in the delay interval is about 0.006$mm$, which can be neglected in the GVLBI delay model (\citealt{Petit+etal+2010}). For SVLBI, the acceleration and the maximum delay are much larger. For example, if the spacecraft has an apogee height of 90,000 $km$, and perigee height of 1,000 $km$, the delay is approximately 0.3$s$, the maximum acceleration will be $8m\cdot s^{-2}$, yielding a 0.36$m$ effect on the baseline in the delay interval. Thus, the spacecraft's acceleration effect should be included in the Space VLBI delay model.

The delay and delay rate solutions inside the VLBI correlator are produced every two-minutes interval (\citealt{Benson+1995}). A series of ten consecutive solutions are windowed into a quantic spline fitting algorithm and resulting fifth-order polynomials are evaluated by the correlator firmware and track the interferometer delays over two minute model intervals (\citealt{Wells+etal+1989}; \citealt{Benson+1995}). For SVLBI, the delay rate is more difficult to address. The spacecraft's maximum speed in an elliptical Earth orbit is about 10$km\cdot s^{-1}$, yielding a delay rate of 33$\mu s\cdot s^{-1}$, which is much larger than the delay rate of GVLBI. With an acceleration close to $8m\cdot s^{-2}$, the SVLBI delay rate may require more accurate polynomial fitting or more frequent model updates to meet certain accuracy.

\section{Delay model for SVLBI}
\label{sect:Obs}
\subsection{Coordinate System Transformation}
A reference system should be chosen so that the physical process of concern can be described as simply as possible. Considering the effect of Earth's revolution in the solar system and the gravitational effect on the signal propagation, the delay model should be described in Barycentric Celestial Reference System (hereafter BCRS), whose time-coordinate is called TCB. Meanwhile, the positions and the velocities of the VLBI stations (both the ground-based VLBI stations and the orbiting space VLBI stations) are given in the Geocentric Celestial Reference System (hereafter GCRS), whose time-coordinate is called TCG. 

The coordinate transformation between BCRS and GCRS given by the XXIV IAU General Assembly is as follows (\citealt{Soffel+etal+2003}):

\begin{equation} t=\int \left (1-\cfrac{U_{ext}(\vec{X_E})}{c^2}-\cfrac{{V_E}^2}{2c^2}\right) \, dT-\cfrac{\vec{V_E}\cdot\vec{X}}{c^2}+O(c^{-4})
 \end{equation}
\begin{equation}
\vec{x}=\vec{R_E}+\cfrac{1}{c^2}\left[U_{ext}\vec{R_E}+\frac{1}{2}(\vec{V_E}\cdot\vec{R_E})\vec{V_E}+(\vec{A_E}\cdot\vec{R_E})\vec{R_E}-\cfrac{1}{2}{R_E}^2\vec{A_E}\right]+O(c^{-4})\end{equation}
Where $U_{ext}(\vec{X_E})=\sum_{P\neq E}\frac{GM_P}{|\vec{X}-\vec{X_P}|}$ is the gravitation potential at the 
geocenter, neglecting the effects of \newpage \noindent the Earth's mass. At picosecond level, only the solar potential need to be included (\citealt{Petit+etal+2010}).$\vec{V_E}$ is the barycentric velocity of geocenter.  $\vec{X_E}$ is the barycentric radius vector of geocenter, $\vec{X}$ is the station's barycentric position vector, while $\vec{x}$ is the GCRS position vector. $T$ refers to the barycentric coordinate time TCB, and $t$ refers to TCG. $c$ is the light speed in the vacuum.  $\vec{A_E}$ is the acceleration of the geocenter.   And $\vec{R_E}=\vec{X}-\vec{X_E}$.

In particular, we assume the proper time of the clock in the orbiting station has been transferred to TCG. The transformation of the proper time of the satellite clock to TCG will not be discussed in this paper, readers can refers to IERS Convention 2010, Chapter 10. 

\subsection{SVLBI Delay Model}

In the barycentric frame, the delay equation is, to a sufficient level of approximation:
\begin{equation}
\vartriangle{T}=T_2-T_1=-\frac{\vec{K}}{c}\left(\vec{X_2(T_2)}-\vec{X_1(T_1)}\right)+\vartriangle T_{grav}
\end{equation}
While $\vec{K}$ is the unit vector from the  barycenmtric to the source in the absence of gravitational or the aberrational bending; $\vec{X_i}$ is the barycentric radius vector of the $i^{th}$ receiver at the TCB time $T_{i}$. $\vartriangle T_{grav}$ is the general relativistic delay.
\begin{figure}[h]
  \centering
  \includegraphics[width=13.5cm, angle=0]{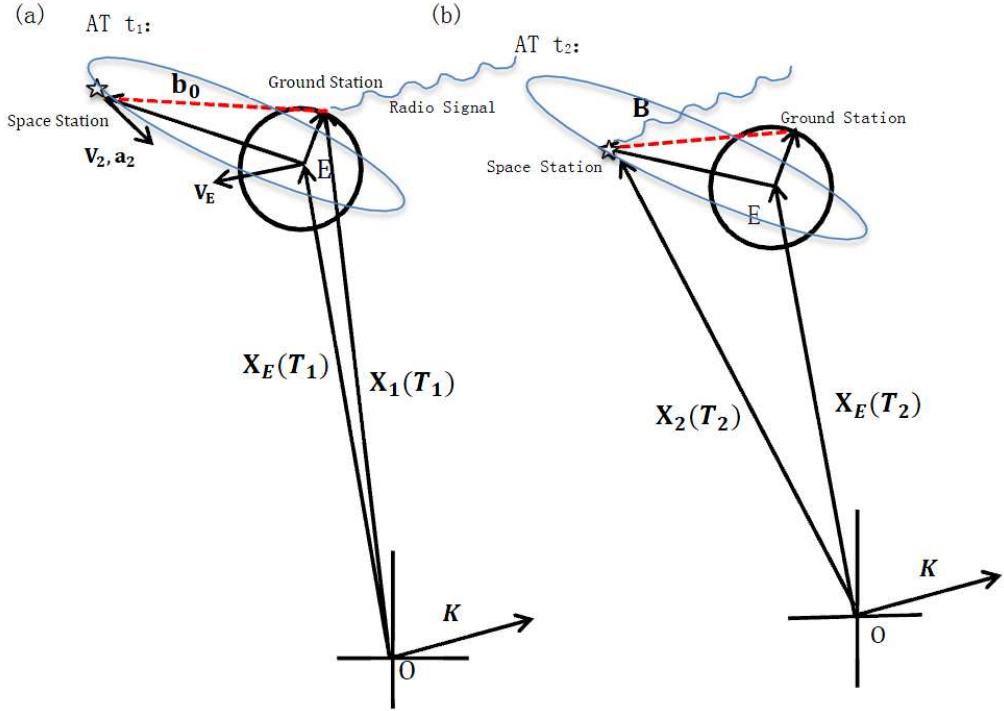}
   \caption{Geometry of the receiving stations in Space VLBI. Figure 1(a) shows the orientation of the signal at the ground station. Figure 1(b) shows the orbit movement of the space station antenna at the time of  reception the signal. And the geocenter of the Earth has moved.} 
  \label{Fig1}
  \end{figure}

The poisition vector of the orbiting station in BCRS can be approximated with coordinate velocity and acceleration as:
\clearpage
\begin{equation}
\vec{X_2(T_2)}=\vec{X_2(T_1)}+\vec{V_E}(T_2-T_1)+\vec{V_2}(T_2-T_1)+\tfrac{1}{2}\vec{a_2}(T_2-T_1)^2
\end{equation}

Where $\vec{V_2}$ is the orbiting velocity of the space station, and $\vec{a_2}$ is its orbting acceleration.
With Eq. (2.4), Eq. (2.3) is re-writtten as
\begin{equation}
\frac{\vec{K}}{2c}\vec{a_2}(T_2-T_1)^2+\left [1+\frac{\vec{K}}{c}(\vec{V_E}+\vec{V_2})\right](T_2-T_1)+\frac{\vec{K}}{c}\vec{B_0}-\vartriangle T_{grav}=0
\end{equation}
With $\vec{B_0}=\vec{X_2(T_1)}-\vec{X_1(T_1)}$.
Eq. (2.5) is a quadratic equation. According to Halley's method (\citealt{Danby+1988}), when a quadratic equation has the following form:
\begin{equation}
\frac{1}{2}Ax^2+Bx+C=0
\end{equation}
Its approximation solution is obtained by
\begin{equation}
x=-\frac{C}{B\left[1-\tfrac{CA}{2B^2}\right]}
\end{equation}
This approximation gives third-order convergence in iterative use to solve a quadratic equation (\citealt{Sekido+etal+2006}). It is quite effective especially when $A \cdot C \ll B^2 $, as is the case here.

The solution of Eq.(2.5) is obtained using Halley's method as:
\begin{equation}
T_2-T_1=\frac{-\frac{\vec{K}}{c}\cdot \vec{B_0}+\vartriangle T_{grav}}{\left[1+\tfrac{\vec{K}}{c}\cdot (\vec{V_E}+\vec{V_2})\right]\left\{{1-\tfrac{(\vec{K}\cdot \vec{a_2})(\vec{K}\cdot \vec{B_0})}{2c^2\left[1+\tfrac{\vec{K}}{c}\cdot (\vec{V_E}+\vec{V_2})\right]^2}}\right\}}
\end{equation}
The approximation error of Halley’s method is evaluated by
\begin{equation}
\delta x\approx \frac{C^3 \cdot A^2}{4B^5} = \frac{\left(\tfrac{\vec{K}}{c}\cdot \vec{B_0}\right)^3\left(\tfrac{\vec{K}}{c}\cdot \vec{a_2}\right)^2}{4\left[1+\tfrac{\vec{K}}{c} \cdot (\vec{V_E}+\vec{V_2})\right]^5}
\end{equation}
The order of $\delta x$ is $\left (\tfrac{\vec{K}}{c}\cdot \vec{a_2}\right)^2 \approx 10^{-16}s$, which can be negligible.

In the case of SVLBI observation, the time interval of signal arrival at two antennas is about $0.1s$ order, the coordinate transformation between BCRS and GCRS can be approximated  up to the order of $(\tfrac{V_E}{c})^2$ for 1 picosecond accuracy, where coordinate velocity of the Earth and the external gravitational potential in the interval of Eq. (2.1) and Eq. (2.2) can be treated as constant:
\begin{equation}
T_2-T_1=\left(1+\frac{U_{ext}}{c^2}+\frac{{V_E}^2}{2c^2}\right)(t_2-t_1)+\frac{\vec{V_E}}{c^2}\cdot \left(\vec{b_0}+\vec{V_2}(t_2-t_1)\right)
\end{equation}
\begin{equation}
\vec{B_0}=\vec{b_0}-\frac{\vec{b_0}}{c^2}U_{ext}+\frac{\vec{b_0}\cdot \vec{V_E}}{2c^2}\vec{V_E}
\end{equation}
Substituting Eq. (2.10) and Eq. (2.11) into Eq.(2.8), the SVLBI delay model can be written as:
\begin{align}
& \vartriangle{t}\!=\!t_2\!-\!t_1 \nonumber \\ 
& \!=\!\tiny{\tfrac{-\tfrac{\vec{K\!\cdot \!\vec{b_0}}}{c}\left(1\!-\!\tfrac{2U_{ext}}{c^2}\!-\!\tfrac{\vec{V_E}\!\cdot\! \vec{V_2}}{c^2}\!-\!\tfrac{{V_E}^2}{2c^2}\right)
\!-\!\tfrac{\vec{K\!\cdot \! \vec{b_0}}}{c}\left(1\!-\!\tfrac{2U_{ext}}{c^2}\!-\!\tfrac{\vec{V_E}\!\cdot\! \vec{V_2}}{c^2}\!-\!\tfrac{{V_E}^2}{2c^2}\right)
\!-\!\tfrac{\vec{V_E}\!\cdot\! \vec{b_0}}{c^2}\left(1\!+\!\tfrac{\vec{K}\!\cdot\! \vec{V_E}}{2c}\right)
\!+\!\tfrac{\vec{K\!\cdot\! \vec{a_2}}}{2c}\left [\tfrac{\vec{K}\!\cdot\!\left(\vec{b_0}\!-\!\tfrac{\vec{b_0}}{c^2}U_{ext}\!+\!\tfrac{\vec{b_0}\!\cdot\! \vec{V_E}}{2c^2}\vec{V_E}\right)}{c}\right ]^2\!+\!\vartriangle T_{grav}}{1\!+\!\tfrac{\vec{K}}{c}\!\cdot\! (\vec{V_E}\!+\!\vec{V_2})}} 
\end{align}
The time derivative of Eq. (2.12) is the delay rate model. 
\clearpage
\section{Model calculation and fitting in VLBI correlator}
In the VLBI correlator, the delay and delay rate models are calculated at two-minute intervals. A series of ten consecutive solutions are windowed into a quintic spline-fitting algorithm (\citealt{Benson+1995}):
\begin{equation}
y=p_1t^n+p_2t^{n-1}+\cdots+p_nt+p_{n+1}, n=5
\end{equation}

And the resulting fifth-order polynomials are evaluated by the correlator firmware and track the interferometer delays over the intermediate two-minute model interval. The next two minutes will be evaluated using another fifth-order polynomials by fitting the next series of solutions. The fitting precision should meet the precision required of the model calculation in VLBI correlator, which has following form (\citealt{Shu+etal+2001}):
\begin{equation}
\vartriangle \tau\le\frac{N}{2B}; \vartriangle \dot{\tau}\le \frac{1}{2T\cdot f}
\end{equation}
Where $B$ is the bandwidth of the base band converter, $N$ is the number of the delay channel, $T$ is the integration time, and $f$ is the observing frequency. 

For example, if $N$ is 32, and $B$ is 16 MHz. The integration time is 4$s$, and the observing frequency of SVLBI can reach 50 GHz, the precision of model calculation is as follows:
\begin{equation}
\vartriangle \tau \le \frac{32}{2\cdot 16MHz}=1\mu s; \vartriangle \dot{\tau} \le \frac{1}{2\cdot 4s\cdot 50GHz}=2.5ps\cdot s^{-1}
\end{equation}
\begin{table}
\bc
\begin{minipage}[]{150mm}
\caption[]{Initial orbit elements of the orbiting telescope\label{tab1}}\end{minipage}
\setlength{\tabcolsep}{1pt}
\small
 \begin{tabular}{cccc}
  \hline\noalign{\smallskip}
Orbit element& & & Value\\
  \hline\noalign{\smallskip}
Semimajor Axis& & &36,978.14km\\
Eccentricity& & &0.79\\
Inclination& & &$28.5^{\circ}$\\
Period& & &19.65h \\
Perigee Altitude& &  &1,200km\\
Apogee Altitude& & &60,000km\\
  \noalign{\smallskip}\hline
\end{tabular}
\ec
\tablecomments{1.1\textwidth}{Two orbitting telescopes are planned to launch in the first step of C-SVLBI, angle between two orbital planes is about $120^{\circ}$. }
\end{table}
\begin{table}
\bc
\begin{minipage}[]{150mm}
\caption[]{Ground-based stations' positions\label{tab2}}\end{minipage}
\setlength{\tabcolsep}{1pt}
\small
 \begin{tabular}{ccccccc}
  \hline\noalign{\smallskip}
 & 65-meter telescope& & &GVLBI-1& & GVLBI-2\\
  \hline\noalign{\smallskip}
Longitude&$121^{\circ}11'59''$E &&&$0^{\circ}$ &&$90^{\circ}$/$180^{\circ}$\\
Latitude& $31^{\circ}05'57''$N&&& $0^{\circ}$&&$0^{\circ}$\\
Altitude(m)& 5&&& 0&&0\\
  \noalign{\smallskip}\hline
\end{tabular}
\ec
\tablecomments{0.6\textwidth}{GVLBI-1 and GVLBI-2 refer to the two ground based telescopes}
\end{table}
In ground VLBI, the maximum delay is about $21ms$, and the delay rate is less than $3\mu s\cdot s^{-1}$. For SVLBI, the maximum delay and fringe rate that must be accommodated are much larger. Take C-SVLBI as an example, the spacecraft has an apogee height of $60,000 km$, and the maximum delay will be approximately $0.2s$. The maximum speed of the spacecraft around the perigee is about $8km\cdot s^{-1}$, corresponding to a delay rate of $26\mu s\cdot s^{-1}$. As shown in Fig. 2.\clearpage
\noindent 
\subsection{Simulation results}
In order to examine the quintic spline fitting’s effectiveness for SVLBI, we show the simulation results in this section.
\begin{figure}[h]
  \centering
  \includegraphics[width=14.5cm, angle=0]{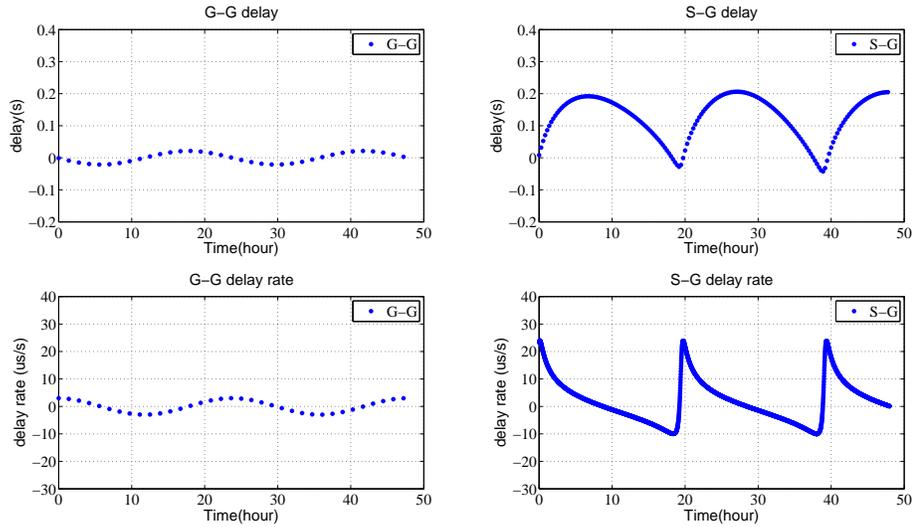}
   \begin{minipage}[]{150mm}
   \caption{The simulation delay and delay rate of GVLBI (G-G) and C-SVLBI (S-G). } 
\end{minipage}
  \label{Fig2}
  \end{figure}
\begin{figure}[h]
  \centering
  \includegraphics[width=14.5cm, angle=0]{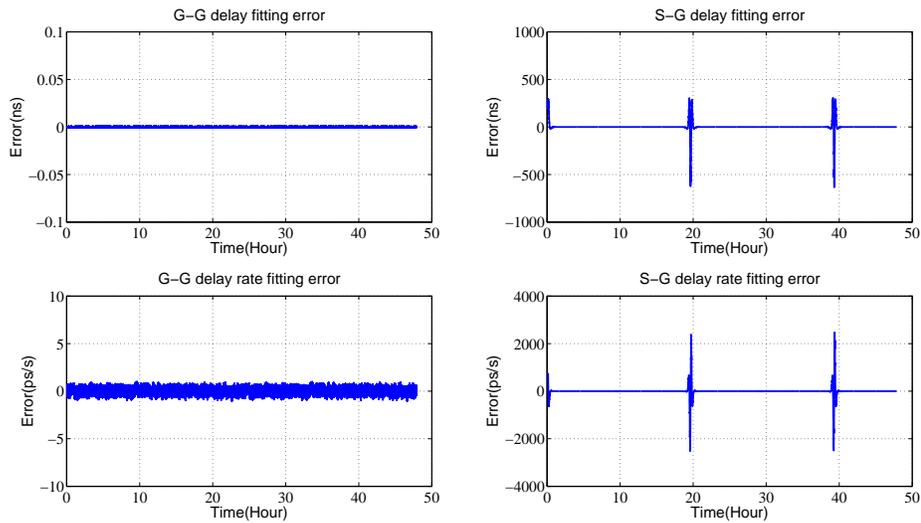}
   \begin{minipage}[]{155mm}
   \caption{The spline fitting error of  delay and delay rate of GVLBI and C-SVLBI. } 
\end{minipage}
  \label{Fig3}
  \end{figure}
\clearpage
\noindent 
The simulation conditions are shown as follows.
The initial orbit elements of the C-SVLBI orbit are shown in Table 1. The simulation time is from Sept 8 2004 at 04:00:00 UT to Sept 10 2004 at 04:00:00 UT. The orbit of C-SVLBI is simulated with the Satellite Tool Kit (STK). We adopt the Shanghai 65-meter radio telescope as the ground station, which observes in conjunction with the orbiting telescope. For GVLBI, in order to get the maximum delay, we adopt two imaginary telescopes, which are located in equator with a difference of 90 degrees in longitude, and 180 degrees difference in longtitude for maximum delay rate, as shown in table 2.

\begin{figure}[htbp]
  \centering
  \includegraphics[width=7.0cm, angle=0]{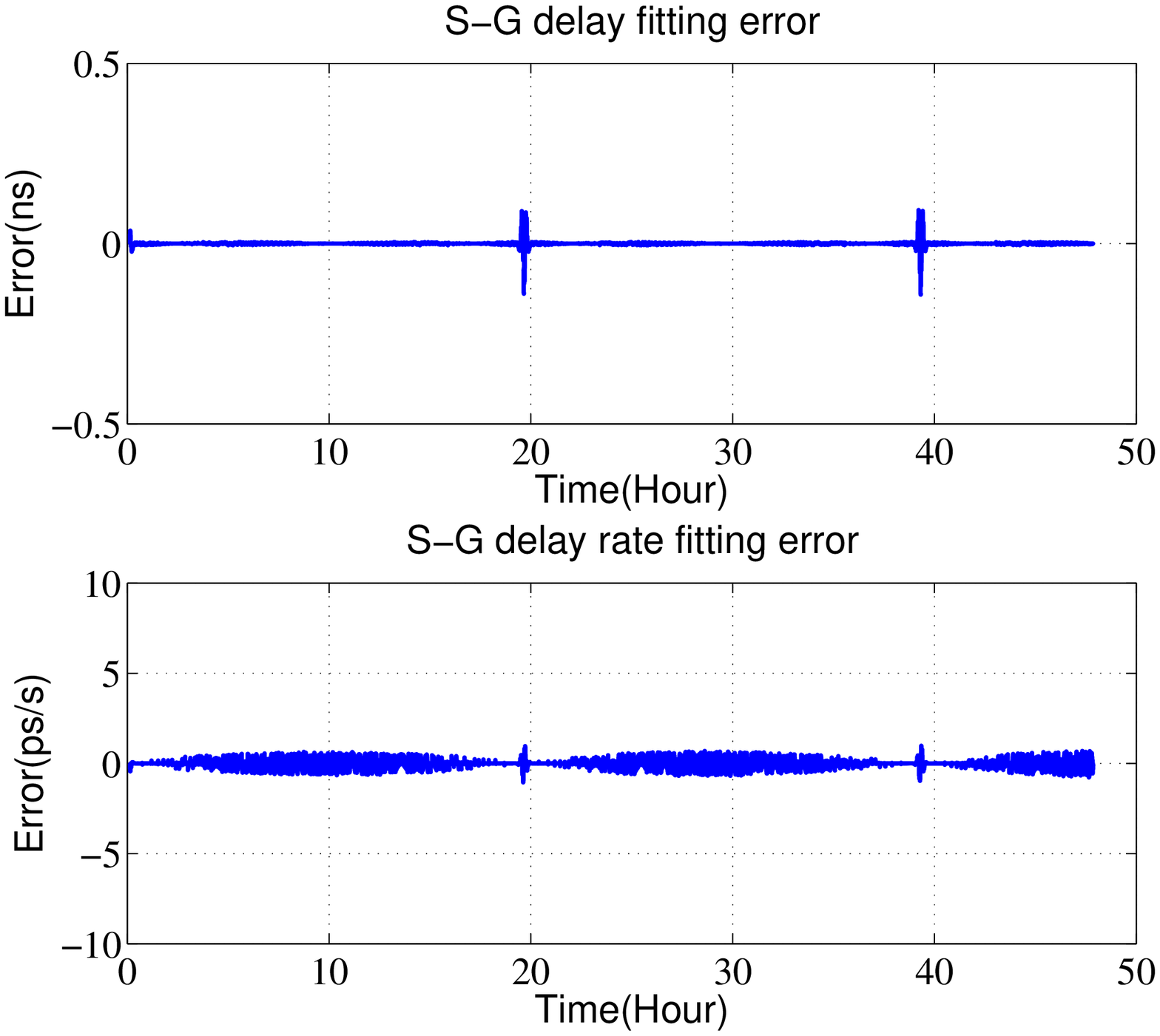}
  \includegraphics[width=7.0cm, angle=0]{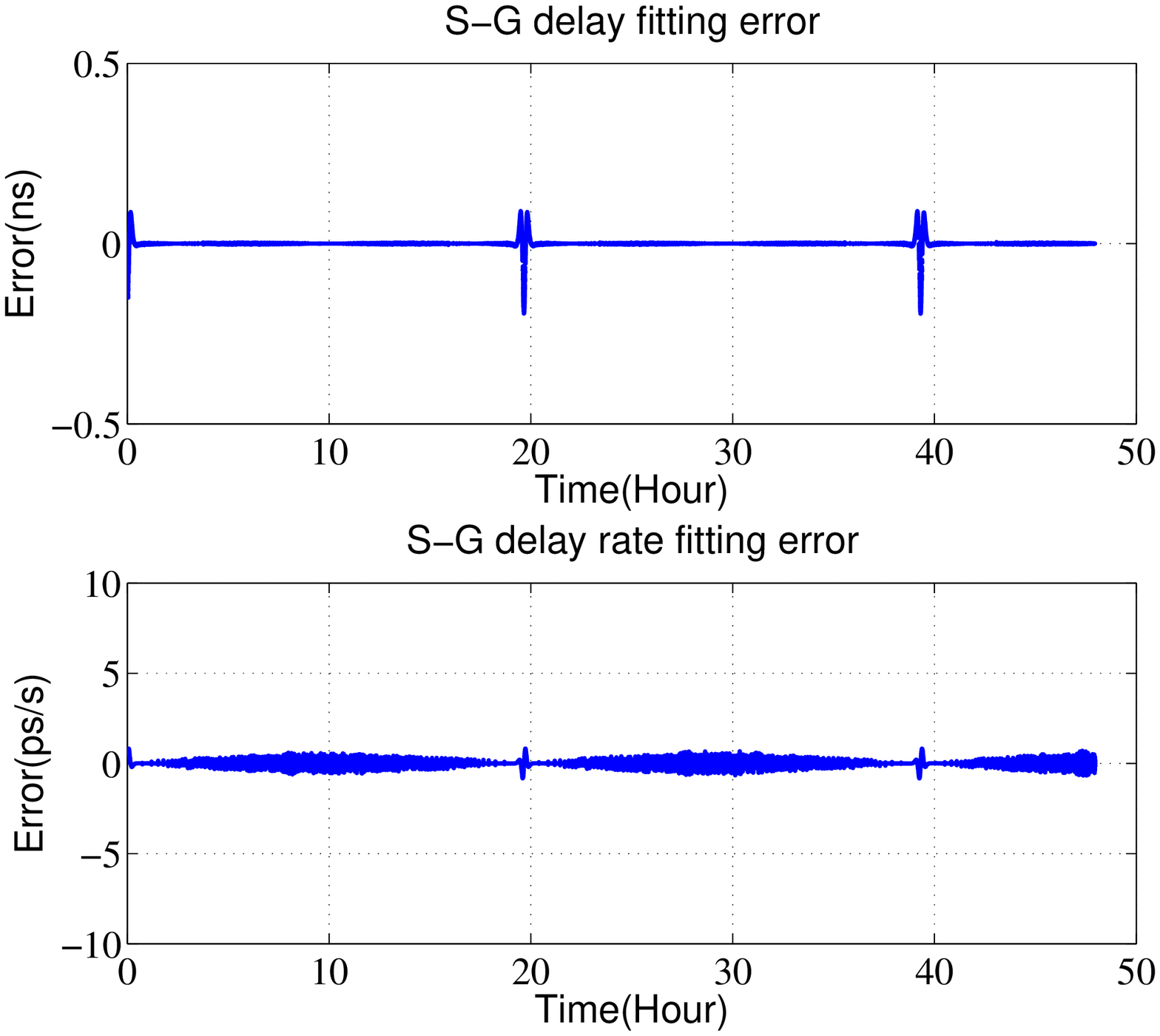}
   \begin{minipage}[]{140mm}
   \caption{The left panel shows the fitting result of 2 minutes interval using 8-order polynomial. The right panel is the fitting error of 30 seconds interval using 5-order polynomial. } 
\end{minipage}
  \label{Fig4}
  \end{figure}
We calculate the SVLBI delay and delay rate by using Eq. (2.12) and its time derivation, and the GVLBI delay and delay rate  by using the formulas given by IERS, the simulation results are shown in Fig. 2. And we use Eq. (3.1) to fit the ten consecutive solutions, then we evaluate the intermediate two-minutes' delay and delay rate. Repeat these steps through the simulation time. The fitting error is obtained by comparing the fitting result with the model calculation result. Compare with the GVLBI, the delay rate of SVLBI is more difficult to address. As shown in Fig.3, around the perigee, the fifth-order fitting error of the delay rate of the SVLBI is about $2ns$, which is much larger than the required precision.

As shown in Fig. 4, if we use eight-order polynomial to fit the delay and delay rate, the fitting result is much better. Another solution is to shorten the time interval of model calculation to 1 minute or less, say $30s$ as shown in the right panel of Fig. 4. Therefor, the SVLBI delay rate requires more accurate polynomial fitting and evaluation, more frequent model updates, especially around the perigee.
\section{Conclusion}
\label{sect:conclusion}
In this paper, a relativistic delay model for SVLBI is derived. The acceleration of the orbiting telescope is much larger than the GVLBI, and should be included in the delay model. The SVLBI delay rates are higher than GVLBI. The delay models inside VLBI correlators are expressed as polynomials in time, the higher delay rate requires more accurate polynomial fitting and evaluation, more frequent model updates.
\clearpage
\bibliographystyle{raa}
\bibliography{bibtex}

\end{document}